\documentclass[%
reprint,
superscriptaddress,
altaffilletter,
 amsmath,amssymb,
 aps,
]{revtex4-1}

\usepackage{natbib}

\usepackage{graphicx}
\usepackage{multirow}
\usepackage{subfigure}
\usepackage{dcolumn}
\usepackage{enumerate}
\usepackage[%
    colorlinks=true,%
    linkcolor=black,%
    citecolor=black,%
    urlcolor=black%
    ]{hyperref}
\usepackage[all]{hypcap} 
\usepackage{color}

\usepackage[mathlines]{lineno}

\begin{document}




\title{\textbf{Study of Supernova $\nu$-Nucleus Coherent Scattering Interactions}}


\author{M.~Biassoni}
\email{Matteo.Biassoni@mib.infn.it}
\affiliation{ INFN - Sezione di Milano-Bicocca, Milano I-20126 - Italy}
\affiliation{Dipartimento di Fisica, Universit\`{a} di Milano-Bicocca, Milano I-20126 - Italy}

\author{C.~Martinez}
\email{carlos@owl.phy.queensu.ca}
\affiliation{Dept.\ of Phys.\ and Astron., Queen's University, Kingston, Ontario K7L 3N6 - Canada}

\date{\today}



\begin{abstract}
Presently, there are several experimental setups dedicated to rare event searches, such as dark matter interactions or double beta decay, in the building or commissioning phases. These experiments often use large mass detectors and have excellent performance in terms of energy resolution, low threshold and extremely low backgrounds. In this paper we show that these setups have the possibility to exploit coherent scattering on nuclei to detect neutrinos from galactic supernova explosions, thus enlarging the number of early detection ``observatories'' available and helping in the collection of valuable data to perform flavour-independent studies of neutrinos' emission spectra.
\end{abstract}





\maketitle

\section{Introduction}
\label{sec:intro}

Neutrinos from core-collapse supernovae are messengers of rich information in both particle physics (neutrino properties, oscillations) and astrophysics (supernova mechanism, very dense matter behaviour). They also constitute, as of today, the only prompt detectable signal of a supernova event, as the technology to detect gravitational waves is still under development and no signal has been detected yet.

Charged current (CC) scattering based experiments such as Super-Kamiokande \cite{ref:superk}, Borexino \cite{ref:borexino} and LVD \cite{ref:lvd} are able to detect incoming electron antineutrinos ($\bar\nu_{e}$) in the supernova energy range with high efficiency by means of the inverse beta decay on free protons (the Cherenkov or scintillation light produced by the positron emitted during the process is the actually detected signal). Electron neutrinos ($\nu_{e}$) and $\nu_{x}$ (sum of $\nu_{\mu}$, $\bar\nu_{\mu}$, $\nu_{\tau}$, $\bar\nu_{\tau}$) can also be detected by these experiments, but the cross sections of the involved processes (CC and neutral current, NC, scattering on electrons) are much smaller. Since the cross sections for neutrino-electron interactions, especially in the MeV energy range, are small, the detector mass has to be overwhelmingly large ($O(1000\,\mathrm{tons})$) to compensate. Furthermore, the detection capability is almost entirely limited to electron neutrinos, while during a supernova explosion neutrinos of all three flavours are supposed to be produced. Hence, the inclusive detection of all neutrino species could provide important and oscillation-independent information about their total emission flux and spectrum.

A very promising but not yet exploited mechanism to detect $\nu_{x}$ is neutrino-nucleus coherent elastic scattering on target nuclei. This process is flavour-blind and, for small enough momentum transfer, the cross section is highly enhanced by the coherent superposition of interaction probabilities for all nucleons within the scattered nucleus. Due to the possibility of detecting all neutrino components and the enhancement of involved cross sections, the expected number of events from a standard supernova turns out to be large enough to make a 1$\,$ton scale detector based on coherent scattering as effective as a 100$\,$ton light water Cherenkov detector. Moreover, the recoil energy of coherently scattered nuclei is correlated to the neutrinos' energy in such a way that some information about the neutrinos spectra, the average temperature for example, can be reconstructed. This is not possible, for instance, with inverse beta decay interactions in scintillation detectors, where the measured deposited energy does not depend on neutrino energy as long as it is above threshold. A large mass coherent scattering detector can therefore be used, in principle, as a thermometer for $\nu_{x}$ emitted by collapsing stars.

Demonstrating the capability of an experiment using coherent elastic scattering to detect supernova neutrinos increases the number of experiments potentially involved in early supernovae detection networks like SNEWS \cite{ref:snews}.

Presently, many experiments for rare events (double beta decay, dark matter search) are in building or commissioning phase. These experiments, which are often based on cryogenic detectors, have in common good energy resolution (hence low threshold capabilities), extremely low background and large masses, and they often use detectors containing high atomic mass elements (Ge, Te, Cd,W). Noble gases (Ar, Xe) and large mass standard scintillating detectors (NaI) are interesting as well. In \autoref{table:experiments} some experiments that could potentially use this technique are reported.
\begin{table}[tb]
\begin{center}
\begin{tabular}{ccc}
&\\
\hline\hline
Experiment & Detector material & Mass [kg] \\
\hline
GERDA (phase II)\cite{ref:gerda} & Ge  & 37.5\footnotemark[1] \\
SuperCDMS (phase B)\cite{ref:cdms}& Ge  & 145\footnotemark[1]  \\
CUORE\cite{ref:cuore}\cite{ref:lowth} & TeO$_{2}$  & 741\footnotemark[2] \\
COBRA\cite{ref:cobra} & CdZnTe & 0.42\\
CRESST\cite{ref:cresst} & CaWO$_{4}$ & 10\\
XENON100\cite{ref:xenon} & Xe & 62\footnotemark[1] \\
WARP\cite{ref:warp} & Ar  & 150\\
DAMA/LIBRA\cite{ref:dama} & NaI  & 250\\
\hline\hline
\end{tabular}
\footnotetext[1]{The experiment feasibility has been demonstrated (project)}
\footnotetext[2]{The experiment is in commissioning phase}
\caption{Experiments that could potentially be able to detect supernova neutrinos through coherent scattering.}
\end{center}
\label{table:experiments}
\end{table}
The purpose of this paper is to show a systematic study of the potential that different materials have as targets for coherent scattering interactions as a function of the target properties and the neutrinos spectra. Materials already used (or planned to be used) in large mass rare event detectors are especially considered.


\section{Theoretical Background}
\label{sec:theoB}

\subsection{Type II SN}\label{type-II-SN}
A core collapse supernova (or type II supernova) is an astronomical phenomenon marking the end of a massive star's life. Models have shown that for stars with masses grater than $\sim$9 solar masses the end of the hydrogen burning phase is followed by a series of predictable cycles of contraction, heating and burning of progressively heavier elements within the star core (which assumes a onion like structure with the heavier element, iron, at the centre). The dynamical stability is granted, in each layer, by the energy produced in the nuclear fusions. However, in the iron core no net energy is produced as no fusion can occur; electron degeneracy pressure is the only force that prevents the core from collapsing. When the Chandrasekhar limit (1.4 M$_{sun}$) is exceeded, gravity becomes stronger than electron degeneracy and the iron core collapses. During the collapse, the temperature and density increase dramatically and two phenomena occur:
\begin{flalign}
&\texttt{Photodisintegration}&\nonumber\\
&(A,Z)+\gamma\,\rightarrow\,(A',Z')+(Z-Z')p+ \\
&\hspace{2.3cm}(A-A'-Z+Z')n\nonumber\\
&\texttt{Inverse Beta Decay}\nonumber\\
&e^{-}+p \rightarrow n+\nu_{e}\label{inverse-beta-decay}\\
&e^{+}+n \rightarrow p+\bar\nu_{e}\nonumber
\end{flalign}
The iron core continues to shrink until the density approaches the nuclear density, and strong forces and neutron degeneracy prevent further collapse. The in-falling matter rebounds, creating an outgoing shockwave that dissociates nuclear matter, losing energy and finally stalling. The interaction between the shockwave and the core generates extreme temperature/density conditions where nucleon bremsstrahlung and pair annihilation take place.
\begin{flalign}
&\texttt{Bremsstrahlung}&\nonumber\\
&(A,Z)+\gamma\,\rightarrow\,\nu_{e,\mu,\tau}+\bar\nu_{e,\mu,\tau}+(A,Z)\label{bremsstrahlung}\\
&\texttt{Pair Annihilation}&\nonumber\\
&e^{+}+e^{-}\,\rightarrow\,\nu_{e,\mu,\tau}+\bar\nu_{e,\mu,\tau}\label{pair-annihilation}
\end{flalign}
These last two (Equations~\ref{bremsstrahlung} and \ref{pair-annihilation}) are Z$_{0}$ mediated neutral current processes.

Numerical simulations \cite{ref:SN-model1} show that the interaction of a small fraction (0.1$\%$) of the neutrinos generated in this phase with the nuclear matter behind the stalled shock should be enough to rise the shock total energy to positive values. Unbounded layers are ejected in the supernova explosion.

The processes in Equations~\ref{inverse-beta-decay}, \ref{bremsstrahlung} and \ref{pair-annihilation} are the mechanisms that generate the neutrino fluxes emitted in the supernova explosion. The individual contributions to the total flux are $\sim$10-20$\%$ from inverse beta decay and $\sim$80-90$\%$ from pair annihilation and bremsstrahlung \cite{ref:SN-model2}.

A simplified model for the neutrino emission is used in literature \cite{ref:gava,ref:vissani2} when the process described in Equations~\ref{bremsstrahlung} and \ref{pair-annihilation} can be considered as the main channels of neutrino production. This approximation is especially valid in the case of detection through coherent scattering. As will be explained in \autoref{subsec:scattering}, coherent scattering is blind to neutrino flavour; processes like the ones in \autoref{pair-annihilation} can thus be considered the main source of the interacting neutrinos.

\begin{figure}[htbp] 
   \centering
   \includegraphics[width=1\columnwidth]{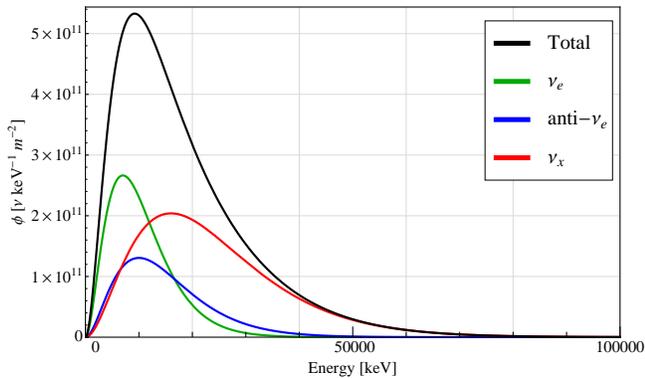} 
   \caption{Boltzmann spectra (at a distance d=8.5$\,$kpc from the source) for the three neutrino families used in the calculations. Green $= \nu_{e}$ (3.5$\,$MeV), blue $= \bar\nu_{e}$ (5$\,$MeV), red $= \nu_{x}$ (8$\,$MeV) and black $=$ total spectrum.}
   \label{fig:spectra}
\end{figure}

The same simplified model predicts the equipartition of the total energy ($\sim3\times10^{53}\,$ergs) among the six neutrino and antineutrino flavours at production. The emission spectra will have different shapes due to the different interaction cross sections, free paths and neutrino spheres' radii for the different species, with resulting different temperatures for $\nu_{e}$, $\bar\nu_{e}$ and $\nu_{x}$, where $\nu_{x}$ are all the remaining neutrino and antineutrino ($\nu_{\mu}$, $\bar\nu_{\mu}$, $\nu_{\tau}$, $\bar\nu_{\tau}$) families (see \autoref{fig:spectra}). Boltzmann spectra with different temperatures are an adequate approximation for our purposes and are used in the literature as well \cite{ref:gava,ref:snews2}.

The spectra in \autoref{fig:spectra} are calculated for a source located at a distance of 8.5$\,$kpc. This is a common assumption in the literature\cite{ref:horowitz,ref:drukier,ref:giomataris}, as it is the distance of the centre of our galaxy. Sometimes 10$\,$kpc is used, as it can be calculated \cite{ref:vissani3,ref:mirizzi} to be the distance with the highest probability of a supernova collapse occurrence.

\subsection{Coherent scattering on target nuclei}
\label{subsec:scattering}

Coherent nuclear elastic scattering is a neutral current weak interaction. From a theoretical point of view, it is the same process of neutrino-nucleon neutral current scattering. If the momentum of the incoming neutrino is small enough, the single nucleon components (protons and neutrons) will not be distinguished and the nucleus will be scattered as a whole. The scattering amplitudes for the different nucleons then coherently sum to give the total cross section. This turns out to be enhanced by a factor of the order of the square of the neutron number compared to that of a single nucleon. The coherent behaviour of the interaction will depend on the actual momentum transfer between the incoming particles (neutrino and nucleus). The higher the momentum transfer, the higher the capacity of the neutrino to distinguish the single components of the nucleus, and hence the smaller the cross section. The nuclear form factor (see \autoref{subsec:target}) is the parameter that accounts for this dependence of the cross section on the momentum transfer.


\section{Experimental Implications}
\label{sec:expimp}

As the aim of this paper is to demonstrate the feasibility of neutrino detection through nuclear coherent scattering, the expected signal for a given detector material, mass, neutrino fluxes and spectra have been calculated. The number of interacting neutrinos depends on:
\begin{itemize}
\item neutrinos' flux
\item neutrinos' energy spectra
\item target nuclear properties
\item number of target nuclei
\end{itemize}
 
In the following analysis, 1$\,$ton of material is considered ($m_{\mbox{\rm\tiny detector}}=1\times10^{6}\,$g) and the corresponding number of target nuclei is calculated, taking into account stoichiometric ratios of the different atomic species
\begin{equation}\label{eq:targetnuclei}
N_{\alpha} = \frac{m_{\mbox{\rm\tiny detector}}}{\sum_{\alpha}A_{\alpha}\eta_{\alpha}} N_{A} \eta_{\alpha}
\end{equation}
where $\alpha$ runs over the nuclear species, $A_{\alpha}$ is the atomic mass (the average over the various isotopes is considered), $\eta_{\alpha}$ is the stoichiometric ratio of the corresponding atom and $N_{A}$ is Avogadro's number.
 
\subsection{Neutrino properties}
As described in \autoref{type-II-SN}, the largest fraction of the energy emitted by a type II supernova ($\sim3\times10^{53}\,$ergs) is carried by neutrinos. The number of neutrinos depends, of course, on the mean energy and the spectral shape. In general, the flux of neutrinos of a given flavour $i = \nu_{e},\bar\nu_{e},\nu_{x}$ at a distance $d$ from the source is
\begin{equation}
\phi_{i}(E)=\frac{\tilde\phi_{i}(E)}{4\pi d^{2}}
\end{equation}
where $\tilde\phi_{i}(E)$ is the emission spectrum at the source. This is assumed to be isotropic.

Though it is obvious, it is still very important to stress the strong dependence of the flux (and consequently the expected number of interacting neutrinos) on the distance at which the supernova occurs. Calculations \cite{ref:vissani3,ref:mirizzi} show that the probability distribution for the distance between the Earth and a supernova does not exclude smaller distances at all.

The spectrum $\tilde\phi_{i}(E)$ can be considered, as a good approximation, to be Boltzmann shaped (\autoref{fig:spectra}),
\begin{equation}\label{eq:spectra}
\tilde\phi_{i}(E) = \frac{N_{i}}{2T_{i}^{3}}E^{2}\exp\!\left(\!-\frac{E}{T_{i}}\right)
\end{equation}
where $T_{i}$  represents the different temperatures ($k_{B}T = 3.5$, $5$ and $8\,$MeV) and $N_{i}$ is the total number of radiated neutrinos. Assuming the equipartition of energy, it is $N_{\nu_{e}} = 3.0\times10^{56}$, $N_{\bar\nu_{e}} = 2.1\times10^{56}$ and $N_{\nu_{x}} = 5.2\times10^{57}$.

Importantly, possible effects of neutrino oscillations have been neglected in this paper. However, this should not introduce any significant error in the signal estimation as the coherent scattering is a flavour-blind process which could, in principle, lead to an inclusive and oscillation-independent detection of all (non-sterile) neutrinos. The spectra of the different neutrino families can then be summed in a total spectrum.
\begin{equation}
\phi_{\mbox{\rm \tiny tot}}(E)=\sum_{i=\nu_{e},\bar\nu_{e},\nu_{x}} \phi_{i}(E)
\end{equation}
Its shape is not analytically defined but strongly depends on the temperatures of the single families. The dependence of the final result (i.e. the amplitude of the detectable signal) on the spectral parameters will be analyzed in \autoref{sec:uncertainties}.

\subsection{Target properties}\label{subsec:target}

\begin{figure}[htbp] 
   \centering
   \includegraphics[width=1\columnwidth]{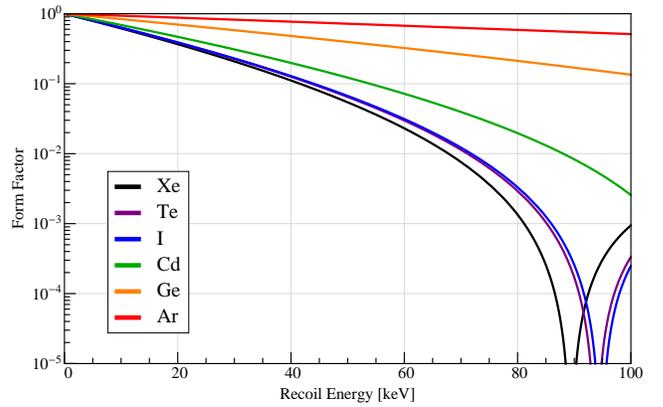} 
   \caption{Form factors for different nuclei.}
   \label{fig:formfactor}
\end{figure}

The cross section for the interaction between incoming neutrinos with energy $E$ and target nuclei via coherent elastic scattering is \cite{ref:horowitz,ref:monroe}
\begin{equation}\label{eq:xsec}
\frac{\mathrm{d}\sigma}{\mathrm{d}\Omega} = \frac{G_{F}^{2}}{4\pi^{2}}E^{2}(1+\cos\theta)\frac{Q_{w}^{2}}{4}F(Q^{2})^{2}
\end{equation}
where $G_{F}$ is the Fermi constant and $\theta$ is the angle between the original and the scattering directions. $Q_{w}$ is the weak charge of the nucleus. This last factor is the one that accounts for the enhancement of the cross section due to the coherent superposition of single-nucleon cross sections
\begin{equation}\label{eq:weakcharge}
Q_{w} = N - (1-4\sin^{2}\Theta_{W})Z
\end{equation}
where $N$ and $Z$ are respectively the number of neutrons and protons within the nucleus and $\sin^{2}\Theta_{W}\approx0.231$, which means that almost only neutrons contribute to the weak charge. The last term in \autoref{eq:xsec} is $F(Q^{2})$, and it is the elastic form factor at momentum transfer $Q$
\begin{equation}\label{eq:transfmom}
Q^{2} = 2E^{2}(1-\cos\theta)
\end{equation}
and represents the distribution of the weak charge within the nucleus. The proton density distribution is often well constrained by measured charge densities, and models exist to calculate the overall form factor. Probably the most complete treatment is in \cite{ref:engel,ref:amanik,ref:lewin} and the form factor used is
\begin{align}\label{eq:formfactor}
F(Q^{2})=&\frac{3\left(\frac{\sin(QR_{0})}{(QR_{0})^{2}}-\frac{\cos(QR_{0})}{QR_{0}}\right)}{QR_{0}} \times \nonumber \\
&\times\exp\left(-\frac{Qs^{2}}{2}\right)
\end{align}
where $R_{0}$ is the nuclear radius defined as 
\begin{align}
R_{0}^{2}&=R^{2} - 5s^{2} \nonumber \\
R=(1.2 \times A^{1/3})\mathrm{fm} \nonumber
\end{align}
and $s$ the nuclear skin thickness ($0.5\,\mathrm{fm}$). \autoref{fig:formfactor} shows the form factor for some of the nuclei considered in this paper. The heavier the nucleus, the more important the correction introduced: the incoming neutrino will no longer coherently see the nucleus as a whole at a smaller momentum transfer if the nucleus is larger, while for smaller nuclei the coherent behaviour lasts until larger values of the momentum transferred in the interaction (corresponding to a larger recoil energy of the nucleus). 

\subsection{Events number calculation}

\begin{figure}[htbp] 
   \centering
   \includegraphics[width=1\columnwidth]{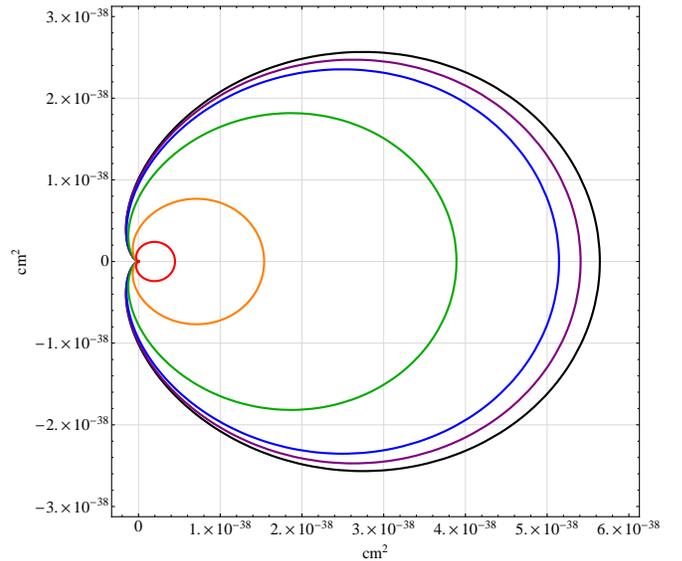} 
   \caption{The polar plot of the coherent scattering differential cross section for different nuclei for a 50$\,$MeV neutrino shows the angular dependence. The colour coding is the same of \autoref{fig:formfactor}.}
   \label{fig:xsec}
\end{figure}

Inserting Equations~\ref{eq:weakcharge} and \ref{eq:formfactor} into \autoref{eq:xsec}, the differential cross section for the cited nuclei can be computed for an interacting neutrino of a generic energy (as an example, \autoref{fig:xsec} represents cross sections for a 50$\,$MeV neutrino).

To obtain the number of interactions within a certain mass of material, the cross section times the neutrino flux has to be integrated. As the cross section depends on the scattering angle, and the scattering angle determines the recoil energy of the scattered nucleus ($2MT=Q^{2}=2E^{2}(1-\cos\theta)$, $M$ is the mass of the recoiling nucleus) through the momentum transfer Q, the events' yield can be obtained through a numerical integration as a function of the kinetic energy of the recoiling nucleus $T$. The analytical form of the integral is
\begin{align}\label{eq:integral}
Y(T)=&\frac{\mathrm{d}N_{\mbox{\rm \tiny events}}}{\mathrm{d}T}=\nonumber\\
=&\sum_{\alpha = \rm{nuclei}} N_{\alpha} \int \!\!\! \int\mathrm{d}\Omega \mathrm{d}E \times \nonumber \\
&\times\frac{\mathrm{d}\sigma_{\alpha}}{\mathrm{d}\Omega}(Q^{2}, F(Q^{2}), Q_{W}^{2}, A) \phi_{\mbox{\rm \tiny tot}}(E) \times \nonumber \\
&\times\delta\left(T-\frac{Q^{2}}{2M_{\alpha}}\right)
\end{align}

where $N_{\alpha}$ is the total number of target nuclei (\autoref{eq:targetnuclei}) and the sum runs over the different nuclear species in the detector material. The three different neutrino species are already summed using $\phi_{\mbox{\rm \tiny tot}}$. In the case the detector contains a single nuclear type (as is the case in Ge or noble gas detectors), the first sum is redundant, while for compound materials (TeO$_{2}$, NaI) the total number of target nuclei for each nucleus depends on the stoichiometric ratio.

\begin{figure}[htbp] 
   \centering
   \includegraphics[width=1\columnwidth]{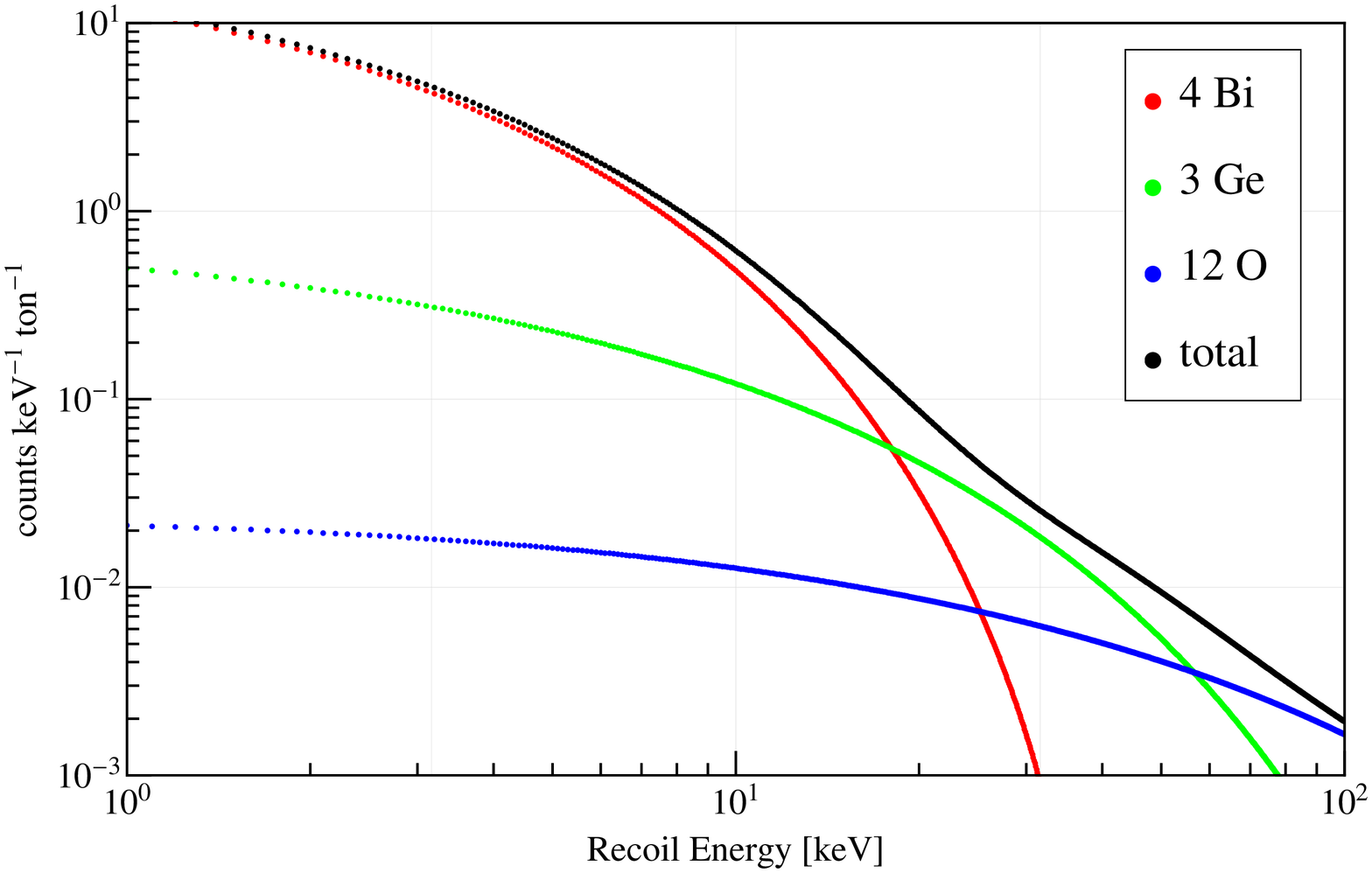} 
   \caption{Events yield for 1-ton of BGO scintillator experiment.}
   \label{fig:YBGO}
\end{figure}

The integration has been performed numerically for each neutrino type independently and for the different nuclei. In \autoref{fig:YBGO} the yield for 1 ton of BGO (Bi$_{4}$Ge$_{3}$O$_{12}$) scintillator is presented as an example.

The different behavior of light and heavy nuclei is evident: the bismuth cross section is enhanced by the large number of neutrons leading to a high number of interactions, but the energy of the recoiling nucleus is always small due to the large mass. Hence, the spectrum of nuclear recoils is steeply exponential shaped. Moreover, high energy recoils are further reduced by the form factor that damps the cross section at high momentum transfer. On the contrary, the number of oxygen recoiling nuclei is much smaller but almost constant as a function of the recoiling energy.

The obvious consequence of the recoil spectrum shape is the importance of the energy threshold and performance of the detector close to it, both in terms of efficiency and background. This is the reason why rare events experiments (that usually have low threshold capabilities and very low and well known background), in particular dark matter and double beta decay bolometric experiments, have the potential for exploiting coherent scattering as a flavour-blind supernova neutrino detection mechanism.
\begin{figure}[h] 
   \centering
   \includegraphics[width=1\columnwidth]{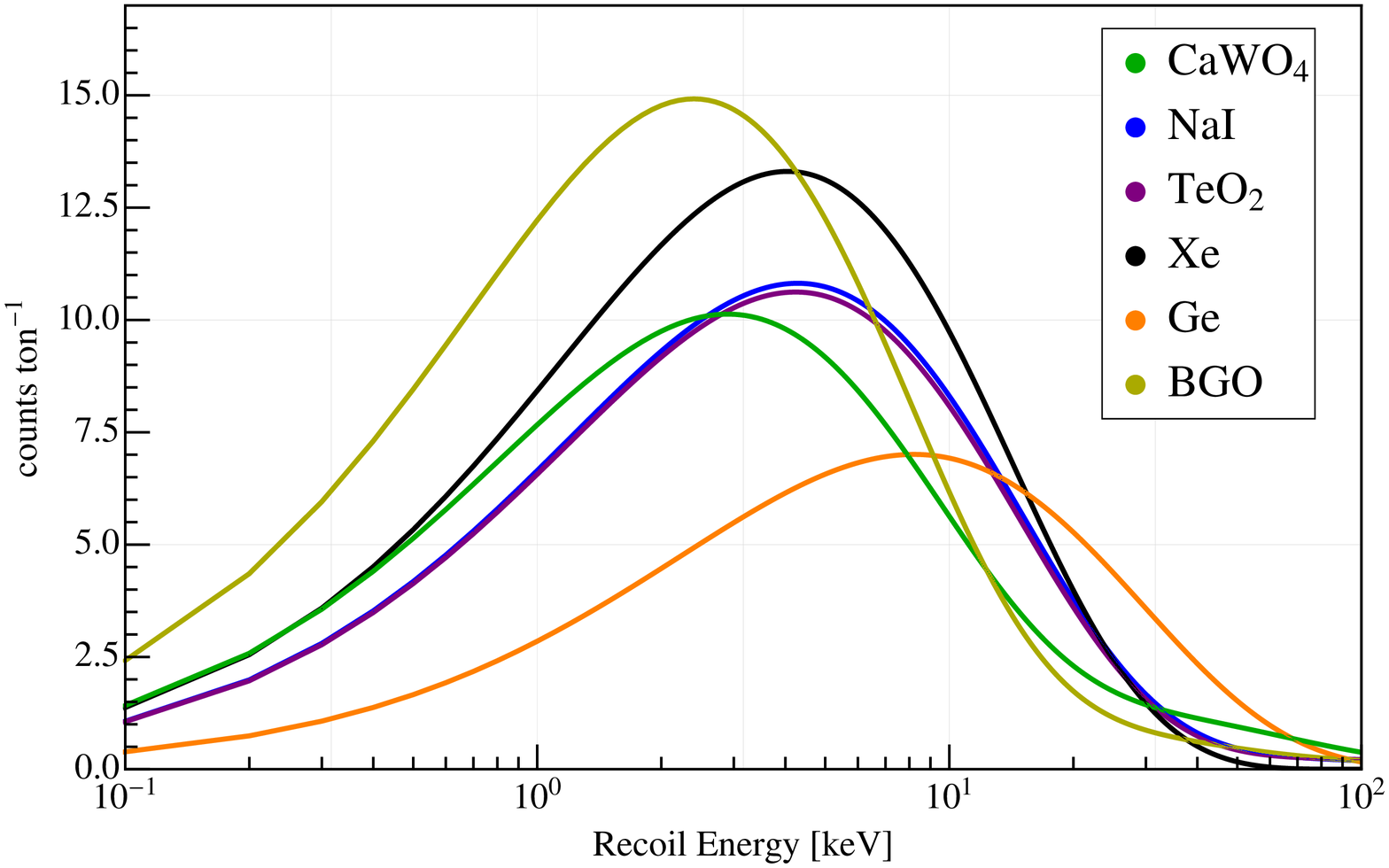} 
   \caption{Events yield for 1-ton of NaI, TeO$_{2}$, Xe, Ge, CaWO$_{4}$ and BGO.}
   \label{fig:Ycomposti}
\end{figure}

\begin{figure}[h] 
   \centering
   \includegraphics[width=1\columnwidth]{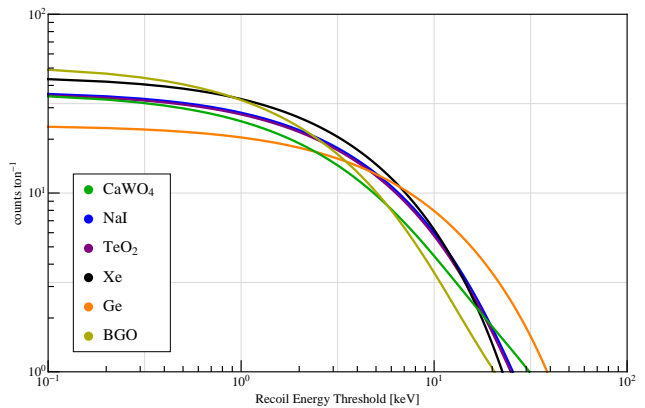} 
   \caption{Number of events above threshold as a function of the energy threshold.}
   \label{fig:intcomposti}
\end{figure}

In \autoref{fig:Ycomposti}, the response function (differential events yield, \autoref{eq:integral}, times the recoil energy) for 1$\,$ton of material is depicted, while in \autoref{fig:intcomposti} the total number of events above threshold as a function of energy threshold is calculated for the same compounds.


\section{Uncertainties propagation}
\label{sec:uncertainties}

The main uncertainties in the results reported in \autoref{sec:expimp} are due to the propagation of uncertainties in the astrophysical models of supernova explosions, hence in the emission spectra of neutrinos. If the equipartition of energy among the species is a widely accepted statement deriving from the universality of the interactions involved in \autoref{bremsstrahlung} and \autoref{pair-annihilation}, parameters such as the total energy of the explosion and the average energy, or the temperature, of the neutrinos spectra are much more uncertain.

A variation in the total energy carried by the neutrinos (assuming a simple luminosity scaling and no change of spectral shape) has the trivial effect of changing linearly the number of neutrinos and hence the number of interactions in a given detector. The effect of a variation of the temperature of the spectra described in \autoref{eq:spectra} must be discussed in more detail. A lower temperature corresponds to a red-shifted emission spectrum. Since the total energy content is the same, but the average energy is smaller, the number of emitted neutrinos is larger but a larger fraction of them will not be able to produce recoiling nuclei above threshold energy. Hence, the signal in the detector drops significantly.

 An increase of the temperature, on the contrary, leads to a higher maximum transferred energy between the incoming neutrinos and the recoiling nuclei. However, since the total number of neutrinos is smaller the signal increase is dumped and tends to saturate. 
 
\begin{figure}[htbp] 
   \centering
   \includegraphics[width=1\columnwidth]{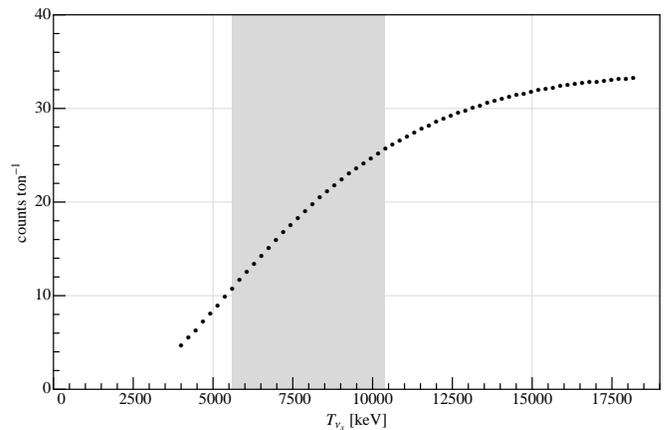} 
   \caption{Number of events in 1 ton of BGO scintillator with 3$\,$keV threshold for different values of neutrinos spectra temperature parameter.}
   \label{fig:tempvariation}
\end{figure}

 The resulting propagation of the uncertainties on the temperature to the signal in the detector is reported in \autoref{fig:tempvariation}. A very large variation of the $\nu_{x}$ temperature is considered while the ratio between the temperatures of the three different species of neutrinos has been kept constant. The result is the number of events in 1$\,$ton of BGO scintillator with an energy threshold of 3$\,$keV. The shadowed region represents a reasonable uncertainty of 30\% on the temperature parameter.


\section{Conclusions}
\label{sec:conclusions}

Reported calculations show that a 1$\,$ton detector with adequate energy threshold (as low as a few keV) has the potential to detect supernova neutrinos through coherent scattering on nuclei. The sensitivity of such an experiment depends on another very important parameter, the background. The number of signals coming from interesting events has to be compared to the number of events coming from the background in order to determine the sensitivity. Rare events experiments usually have very low backgrounds and many of the detectors used in this field are able to distinguish heavy particle (nuclear) recoils from electron recoils through simultaneous measurement of different energy channels, often becoming zero-background experiments. As recently suggested in \cite{ref:articolofiorini}, these experiments' capabilities could not be limited to supernovae neutrino detection, but extend to low energy neutrino measurements in a wider framework including short baseline oscillation studies with very intense sources.


\section*{Acknowledgements}
We owe many thanks to Francesco Vissani for all his help and the many fruitful discussions we had. We would also like to thank several people at the Laboratori Nazionali del Gran Sasso and the Universit\`{a} di Milano-Bicocca, in particular Oliviero Cremonesi, Carlo Bucci and Maura Pavan for their comments and guidance.







\bibliography{\jobname}


\end{document}